\documentclass{llncs}

\usepackage{amsfonts, amsmath}
\usepackage{graphicx,color}
\usepackage{listings}
\usepackage{algorithm}
\usepackage[linesnumbered,ruled,algo2e]{algorithm2e}
\usepackage[noend]{algpseudocode}
\usepackage{wrapfig}
\usepackage{tikz}
\usepackage{url}
\usepackage{wrapfig}
\usepackage{adjustbox}
\usepackage{caption}
\usepackage{subcaption}
\usepackage{marginnote}
\usepackage{mathtools}
\usetikzlibrary{shapes,arrows}

\algrenewcommand\algorithmicrequire{\textbf{Precondition:}}
\algrenewcommand\algorithmicensure{\textbf{Postcondition:}}

\newcommand\f\varphi
\newcommand\Id\varepsilon

\newcommand\lin{\textsf{lin}}
\renewcommand\exp{\textsf{exp}}

\newcommand\dvar{\underline{l}}

\usepackage{setspace}
\usepackage[colorinlistoftodos,prependcaption,textsize=tiny]{todonotes}

\newcommand{\R}{\mathbb{R}}

\newcommand\donotshow[1]{}

\newcommand{\rs}{\textsf{rs}}

\newcommand{\bsamp}{B}
\newcommand{\ParamSet}{\mathcal{P}}

\newcommand{\swsem}[1]{\mathcal{S}(#1)}
\newcommand{\unitsphere}{\partial D}
\newcommand{\specp}{\varphi_{\emph{pulse}}}
\newcommand{\spech}{\varphi_{\emph{heart}}}

\newcommand{\stlspec}{\psi_{\emph{stab}}}

\title{Sampling of Shape Expressions}
\author{Nicolas Basset~\inst{1}, Thao Dang~\inst{1}, Felix Gigler~\inst{2}, Cristinel Mateis~\inst{2}, Dejan Ni\v{c}kovi\'{c}~\inst{2}}

\institute{Université Grenoble-Alpes, Grenoble, France \and AIT Austrian Institute of Technology GmbH, Vienna, Austria}

\begin{document}
\maketitle



Cyber-physical systems (CPS) are 
increasingly becoming driven by data, using multiple types of sensors 
to capture huge amounts of data. Extraction and characterization of useful 
information from big streams of data is a challenging problem. Shape expressions facilitate 
formal specification of rich temporal patterns encountered in time series as well as in behaviors of CPS. 
In this paper, we introduce a method for systematically 
sampling shape 
expressions. The proposed approach combines methods for uniform sampling 
of automata (for exploring qualitative shapes) with hit-and-run Monte Carlo sampling procedures (for 
exploring multi-dimensional parameter spaces defined by sets of 
possibly non-linear constraints). We study and implement several possible solutions and evaluate them in the context 
of visualisation and testing applications.


\section{Introduction}
\label{sec:intro}
%
Cyber Physical Systems (CPS) are 
becoming increasingly driven by data. Today, typical CPS applications 
employ a large variety of sensors to measure, gather and analyse 
huge amounts of data that capture multiple aspects 
of the system, its physical characteristics, 
its internal state and operation, and its environment. 
The availability of this incredible amount of data enables 
new opportunities to improve the 
system operation. It requires analysing raw data, extracting
and characterizing useful information from it 
-- an extremely challenging task that has been traditionally
studied by the time-series analysis in the 
machine learning community~\cite{DBLP:reference/dmkdh/RatanamahatanaLGKVD10}.

{\em Shape Expressions} (SE) is a specification language that facilitates 
the characterization of complex shapes encountered in time series~\cite{DBLP:conf/rv/NickovicQFMD19}. It is an {\em explainable} and {\em interpretable} formalism that 
allows to describe {\em mean} behaviors in presence of 
noise. SEs consist of regular expressions over atomic parametric shapes augmented with constraints on these parameters. 
SE specifications can be designed using the domain knowledge or can be automatically mined from raw time
series. These specifications can represent an important artefact 
in documenting expected or observed behavioral patterns, in detecting anomalies and in understanding 
the overall system behavior.

\begin{figure}
        \centering
    \includegraphics[width=\textwidth]{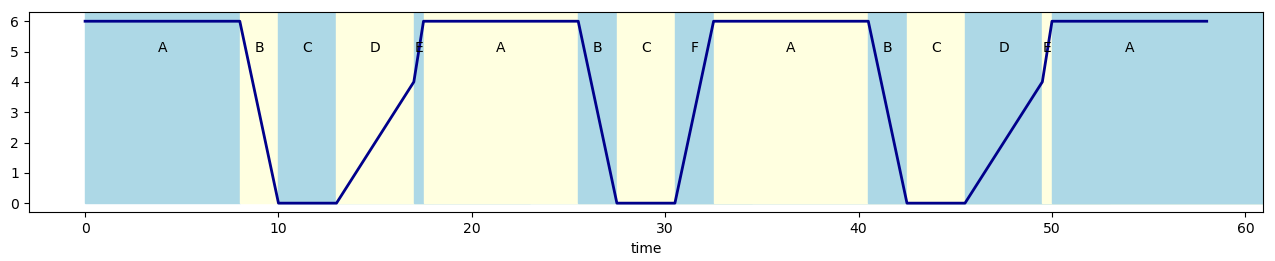}
\caption{Illustrating example - a pulse train.}
\label{fig:pulse}
\end{figure}

We illustrate the specification of sequential behaviors 
in SE with a {\em pulse train} example, depicted in Figure~\ref{fig:pulse}. A pulse train is a sequence of 
{\em pulses}, where each pulse is characterized by 
a sequence of several segments: (A) a constant
segment at some value in $(4, 10)$ with duration in $(6, 10)$, 
(B) a linearly decreasing segment with slope 
in $(-10, -1)$ and duration in $(1, 4)$, 
(C) another constant segment at $0$ with 
duration in $(2, 4)$, and two consecutive 
linearly increasing segments, (D) with slope 
in $(1,2)$ and duration in $(0.5, 2)$, and (E) with slope in
$(5, 900)$ and duration in $(0.01, 2)$.
We finally also allow the concatenation of segments (D) 
and (E) to be replaced by a single segment (F), 
a linearly increasing
segment with slope in $(1, 5)$ and duration 
in $(2,10)$.  
We observe that the above specification 
uses parametric shapes, where
the parameters are possibly constrained, and describes perfect shapes without
accounting for noise.
We express the pulse train specification in SE as 
follows:
$$
\begin{array}{lcl}
\specp & = & (A\cdot B \cdot C \cdot (D\cdot E \cup F))^{+} \cdot A~:~\gamma, 
\text{where} \\ 
\end{array}
$$
$$
\begin{array}{lclclcll}
A & = & \lin_x(a_{1},b_{1},d_{1}) &\;\;\;\;\;\; & \gamma & = & a_1 = 0 \wedge b_{1} \in (4,10) \wedge d_{1} \in (6,10) &\wedge\\ 
B & = & \lin_x(a_{2},b_{2},d_{2}) & & & & a_2 \in (-10,-1) \wedge d_{2} \in (1,4) &\wedge\\
C & = & \lin_x(a_{3},b_{3},d_{3}) &&&& a_3 = 0 \wedge b_{3} = 0 \wedge d_{3} \in (2,4) &\wedge\\
D & = & \lin_x(a_{4},b_{4},d_{4}) &&&&a_4 \in (1,2) \wedge d_{4} \in (0.5,2) & \wedge \\
E & = & \lin_x(a_{5},b_{5},d_{5}) &&&&a_5 \in (5,900) \wedge d_{5} \in (0.01,2) &\wedge\\
F & = & \lin_x(a_{6},b_{6},d_{6}) &&&& a_6 \in (1,5) \wedge d_{6} \in (2,10) &\wedge \; \gamma' \\
\end{array}
$$
\noindent where $(a, b, d)$ denote the slope, the relative offset and 
the duration of a segment, respectively. 
In order to make sure that every segment in the sequence starts where the previous segment ends, we set $\gamma'$ to the following constraint:
$$
\small
\forall i \in [1,4], b_{i+1} =  a_{i}d_{i} + b_{i} \wedge
b_{6} =  a_{3}d_{3} + b_{3} \wedge 
b_{1} = a_{5}d_{5} + b_5 = a_{6}d_{6} + b_6.
$$

We first observe that despite an explainable syntax, the exact meaning of the specification 
$\varphi$ may not be self-evident. Hence, a graphical visualisation may 
help the designer to interpret the specification. This observation is also valid for other specification formalisms such as Signal Temporal Logic STL~\cite{DBLP:conf/formats/MalerN04}, which motivates the work on automatically generating signals satisfying STL \cite{PrabhakarLK18}.  

We also note that SE specifications can be a valuable asset 
in the testing activities.
Indeed, cyber-physical systems rarely operate in unconstrained environments 
and their testing or falsification should ideally 
reflect realistic environment  
profiles. An environment (input) profile 
represents a possibly infinite set of 
input signals subject to various constraints. It is typically obtained from 
the domain knowledge of the 
engineer, but also more recently from mining 
the recorded environment behavior. As an example, the pulses defined and constrained by 
$\varphi$ can be used as inputs from the environment to stimulate a system under test. In addition, shape expressions provide an abstraction that retains essential features of a complex dynamical component, which generates inputs to another component, so that the whole system can be tested in a compositional way. We will illustrate, via a control system case study, the usefulness of SE in specifying constrained behavior spaces that arise in CPS falsification and testing. 
The importance of constraining the input space was already observed in early work on hybrid systems falsification/testing ~\cite{DBLP:conf/hybrid/NghiemSFIGP10}. More generally, it is closely related to constrained random verification initially developed for electronic designs (see for example~\cite{ConstrainedVerifBook2006,KitchenKuehlmann2007}), where constraints are introduced not only to ensure that the stimuli are valid or realistic (so as to avoid false negative testing results), but also to guide test executions towards critical behaviors. When there is no information about the inputs that falsify the property, it is of interest to have a more principled exploration 
of the input space. 
For these reasons, we focus on the problem of automated systematic random generation of behaviors defined by SE specifications, useful for both testing/falsification and  specification visualizing. 

The pulse train example exposes many challenges 
associated to the automated 
generation of behaviors from shape expressions. 
First, there is the problem of 
potentially high number of dimensions. Every line segment 
in $\varphi$ is uniquely characterized by its (slope, relative offset, duration) parameters. After removing parameters that do not change in the pulse example, 
the specification still has $11$ dimensions to be explored. 
Second, the requirement that 
every segment starts where the previous 
one ends introduces constraints, which are both non-linear and 
``thin''. Finally, the specification can include structural 
non-deterministic choices, in the form of union and 
Kleene star operations that need to be taken into account when sampling.

In this paper, we propose a procedure for systematic coverage-based generation of behaviors 
from SE specifications. We propose a two-stage sampling procedure: (1)  we first choose a random word accepted by the specification with uniform distribution using the Boltzmann sampling technique \cite{boltzmann-sampling}; 
(2) in the second stage, we 
choose a random parameter valuation that satisfies the constraints using a 
{\em hit-and-run}~\cite{hitrun} sampling method. 
Hit-and-run is a Markov Chain Monte Carlo (MCMC) sampling method that iteratively 
generates a sequence of points that satisfy a constraint 
by taking steps of (uniformly) random length in 
(uniformly) random directions. Hit-and-run has several desired 
characteristics with respect to the above identified 
challenges: (1) it can be used to sample 
arbitrary open sets in $\R^{n}$, (2) it asymptotically converges to uniform sampling, and (3) the complexity of the
procedure is polynomial in the number of dimensions when the constraints are convex. 
We implement our procedure and evaluate it on multiple case studies, 
including visualization of
ECG heart-beat specifications and test generation of inputs 
constrained by SE in an avionic application.


\section{Shape Expressions}
\label{sec:back}
In this section, we present {\em shape expressions} (SE), a formal specification 
language introduced in~\cite{DBLP:conf/rv/NickovicQFMD19}.
An SE is a regular expression over parameterized {\em atomic shapes}, such as line, exponential and sinsuoid shapes. 

Let $P = (p_1,...,p_n)$ be a set of {\em parameter variables}. A \emph{parameter valuation} $v$ is defined as the mapping
$v: P \rightarrow \mathbb{R}$. We denote $(v(p_1),...,v(p_n))$ as $v(P)$.
A \emph{constraint} $\gamma$ is a Boolean combination of inequalities over P. 
A valuation $v$ {\em satisfies} $\gamma$ (denoted by $v \models \gamma$) iff replacing all variables 
$p \in P$ appearing in $\gamma$ with $v(p)$ makes $\gamma$ evaluate to true. 
The set of all constraints over P is denoted by $\Gamma(P)$.
For a constraint $\gamma\in \Gamma(P)$, we denote by $\ParamSet(\gamma)$ the set of parameter valuations $v$ that satisfies $\gamma$, 
that is $\ParamSet(\gamma)=\{v \mid v \models \gamma\}$. 
Further, we assume that every constraint $\gamma$ involved in this paper yields a set $\ParamSet(\gamma)$ 
whose volume is well-defined and non-null.  

Let X be a set of real-valued signal variables and [0,d) denote a time interval for some $d\in \mathbb{R}_{\geq 0}$.
A {\em signal} $w$ is a function $w: X \times [0,d) \rightarrow \mathbb{R}$  
that maps variables in $X$ at times in $[0,d)$ to real values. The {\em length} of $w$ is denoted by $|w|=d$.
Signals with an empty time domain are permitted and result in the unique empty signal. 
Let 
$w_1 : X \times [0,d_1) \rightarrow \R$ 
and
$w_2 : X \times [0,d_2) \rightarrow \R$
be signals.
Their concatenation denoted by $w_1 \cdot w_2 \equiv w$ is then defined as
\begin{equation*}
    w : X \times [0,d_1 + d_2) \rightarrow \R,~ w(x,t) = 
    \begin{cases}
        w_1(x,t) & \text{if } t \in [0,d_1) \\
        w_2(x,t-d_1) & \text{if } t \in [d_1, d_1 + d_2)
    \end{cases} 
\end{equation*}

While shape expressions can capture specifications 
matched by noisy signals~\cite{DBLP:conf/rv/NickovicQFMD19}, 
in this paper we focus on ideal signals which do not have noise.

 
\subsection{Shape Expressions}

Let $X$ be a set of real-valued signal variables and $P$ a set of real-valued parameters.
%
A shape expression is an expression $\phi = \psi:\gamma$, where $\psi$ is given by the grammar
\begin{eqnarray*}
\psi &::= &\epsilon \mid f_{x}(p_{1}, \ldots, p_{n}, \dvar) \mid \psi_1 \cup \psi_2 \mid \psi_1 \cdot \psi_2 \mid \psi^*,
\end{eqnarray*}
\noindent$x \in X$, $\{p_{1}, \ldots, p_{n}\} \subseteq P$, $f_{x}(p_{1}, \ldots, p_{n}, \dvar)$ is an atomic shape defined by the $n$-ary function $f$ over the variable $x$ with the duration 
$\dvar$ and parameters $\{p_{1}, \ldots, p_{n}\}$ and $\gamma \in \Gamma(P)$. The symbol $\epsilon$ denotes the {\em empty word}, the operators $\cup$, 
$\cdot$ and $*$ denote the classical regular expressions 
{\em union}, 
{\em concatenation} and {\em Kleene star} 
respectively. Hence $\psi$ is a classical regular expression over atomic shapes. 
The set of shape expressions over $P$ and $X$ is denoted $\Phi(P,X)$. 

The regular expression $\psi$ captures the qualitative aspect of the specification, 
while $\gamma$ contains the quantitative information of the shape expression, meaning the global constraints imposed on its parameters. The semantics of
$\psi$ follows the classical semantics of regular expressions, except for the case of an atomic shape. We say that the segment of $w$ defined by its starting and ending times $t$ and $t'$ 
matches $f_{x}(p_{1}, \ldots, p_{n} ,\dvar)$ if $\dvar = t'-t$ and there exists valuation $v$ over 
the parameters $p_{1}, \ldots, p_{n}$ such that $w_{x}[t, t'] = f(v(p_{1}), \ldots, v(p_{n}))[0, t' - t]$.
A sequence of atomic shapes that satisfies $\psi$ will be called a \emph{shape word} of $\f$ and the set of such shape words is denoted by $\swsem{\f}$.

A signal $w$ matches a shape expressions if (1) $w$ can be segmented 
into a sequence of atomic shapes $u$ satisfying the SE's regular expression, and (2) there exists a parameter valuation
under which for 
each atomic shape in the sequence the observed segment and the
instantiated atomic shape are the same.
We also say that the signal $w$ matches $u$.

\section{Sampling Shape Expressions}
\label{sec:sampling}
In this section, we propose a procedure for sampling signals that match a given shape expression $\f=\psi:\gamma$. 
%
Our sampling procedure consists of two samplers: a word sampler and a point sampler. The word sampler selects shape words
$u$ that satisfy the given regular expression $\psi$
and the point sampler selects parameter valuations that satisfy the constraint $\gamma$. The point sampler is based on a hit-and-run algorithm that iteratively updates candidate samples.  Before describing these samplers in detail we state the properties they guarantee.
\begin{theorem}\label{theorem:prop}
\begin{enumerate} 
\item The word sampler 
(1a) samples shape words of different length;
(1b) the mean length of output word can be controlled; 
(1c) it is uniform for every possible length of output sample, {\em i.e.} every two words of the same length have the same probability to be chosen.
\item  
Let $v_k\in \ParamSet(\gamma)$ denote the parameter valuation satisfying the constraint $\gamma$ sampled by the point sampler after $k$ steps. 
Then, for any set $U \subseteq \ParamSet(\gamma)$ and any given value $\epsilon > 0$, there exists $K_{\epsilon}$ such that the probability that $v_{k}$ is in $U$ satisfies:
$$\forall k\geq K_{\epsilon} \, \left|Prob(v_{k} \in U) - \frac{vol(U)}{vol(\ParamSet(\gamma))}\right| < \epsilon.$$
\end{enumerate}
\end{theorem}

The statements (1a-c) show the probability of selecting uniformly a shape word $u$ with a given mean length. 
The statement (2) shows that the point sampler can sample a parameter valuation inside
$\ParamSet(\gamma)$ with the distribution converging towards the uniform distribution. We recall here that each valuation of $\ParamSet(\gamma)$ leads to a noiseless signal so exploring well the set $\ParamSet(\gamma)$ enables one to explore well the noiseless signals matching the selected shape word. 
In the following we provide a description of the two samplers which includes the proof sketches of the above theorem. More concretely, the statements (1) are a direct corollary of  Proposition~\ref{propsition:botzmann}, and the statement (2) is proven by the asymptotic uniformity of hit-and-run, as discussed in the end of Section~\ref{sec:params}.
Note that instead of using the word sampler one can choose a shape word and run the point sampler to explore the set of signals that matches it.

\subsection{Sampling shape words}
\label{sec:paths}
We now explain our word sampler that randomly samples 
shape words from a shape expression following techniques known as Boltzmann sampling \cite{boltzmann-sampling}. Here we give a self-contained presentation of Boltzmann sampler for regular expressions. 
Before defining the Boltzmann sampling algorithm, 
we need the notion of {\em generating functions}. A generating function $g_\psi$ for a regular expression $\psi$ is defined by the series  $g_\psi(z)
=\sum_{w\in \swsem{\f}}z^{|w|}$. 
We denote by $Rconv(g_\psi)$ the convergence radius of $g_\psi$, that is  $Rconv(g_\psi)=\sup\{r\geq 0~|~\forall z<r,~g_\psi(z)<+\infty\}$. 

\begin{proposition}
The generating function of a regular expression is a rational function. It can be computed inductively using  
Algorithm \ref{alg:genfun} provided the regular expression is unambiguous\footnote{
An ambiguous regular expression 
can be disambiguated by first translating the original 
expression to a deterministic finite automaton (DFA), 
and then translate the DFA into an unambiguous regular
expression.}.
\end{proposition}
\begin{proposition}\label{propsition:botzmann}
Let $z<Rconv(g)$. 
The function $p_{\psi,z}$ from $\swsem{\f}$ to $\R_{\geq 0}$ such that $\displaystyle{p_{\psi,z}(u)=\frac{z^{|u|}}{g_{\psi}(z)}}$
is a probability distribution ($\sum_{u\in \swsem{\phi}}p_{\psi,z}(u)=1$).
The mean length $N_z=\sum_{u\in \swsem{\phi}}p_{\psi,z}(u)|u|$ of shape words sampled according to this distribution is a rational function of $z$ that can be computed from $g_{\psi}(z)$ and its derivative $g'_{\psi}(z)$ as follows:
$N_z=z\frac{g'_{\psi}(z)}{g_{\psi}(z)}$. 
 Algorithm~\ref{alg:bsamp} is a \emph{Boltzmann sampler} of parameter $z$, {\em i.e.} it samples a shape word distributed according to $p_{\psi,z}$. 
\end{proposition}
\begin{remark}\label{remaek:root} To tune parameter $z$ to have mean length $N$ it suffices  to find the unique root of the polynomial $Nq(z)-p(z)$ between $0$ and $Rconv(g_{\psi})$ where $p(z)$ and $q(z)$ are polynomials such that $N_z=p(z)/q(z)$.
\end{remark}

Proposition~\ref{propsition:botzmann} proves the statement (1) of Theorem~\ref{theorem:prop}, since any Boltzmann sampler guarantees all three properties (1a, 1b, 1c) in this statement.

\noindent
\begin{small}
\begin{minipage}[t]{.5\textwidth}
\null
\begin{algorithm2e}[H]
\begin{algorithmic}
\State{Switch case}
\State{1. $\psi=\epsilon$ : \Return $1$} 
\State{2. $\psi=a$ : \Return $z$} 
\State{3. $\psi=\psi_1 \cdot \psi_2$ : \Return $g_{\psi_1}g_{\psi_2}$}
\State{4. $\psi=\psi_1\cup\psi_2$ : \Return $g_{\psi_1}+g_{\psi_2}$}
\State{5. $\psi=\psi_0^*$ : \Return\footnote{$g_{\psi}$ satisfies the equation  $g_{\psi}=1+g_{\psi_0}g_{\psi}.$} $1/(1-g_{\psi_0})$}
\caption{Generating function $g_{\psi}$\label{alg:genfun}}
\end{algorithmic}
\end{algorithm2e}
\end{minipage}
\begin{minipage}[t]{.5\textwidth}
\null
\begin{algorithm2e}[H]
\begin{algorithmic}
\State{Switch case}
\State{1. $\psi=\epsilon$ : \Return ($\epsilon$)} 
\State{2. $\psi=a$ : \Return ($a$)} 
\State{3. $\psi=\psi_1 \cdot \psi_2$ :  \Return $\bsamp(\psi_1)\cdot\bsamp(\psi_2)$}
\State{4. $\psi=\psi_1\cup\psi_2$ : \Return $\bsamp(\psi_1)$ with probability $g_{\psi_1}(z)/g_\psi(z)$ and $\bsamp(\psi_2)$} otherwise
\State{5. $\psi=\psi_0^*$ :
\Return $\epsilon$ with probability $1/g_{\psi(z)}$ 
 and $\bsamp(\psi_0)\cdot\bsamp(\psi)$ otherwise}
\caption{Boltzmann sampler $\bsamp(\f)$\label{alg:bsamp}}
\end{algorithmic}
\end{algorithm2e}
\end{minipage}
\end{small}

\begin{example}
The shape expression $\psi:\gamma$ specifying the 
pulse train from Section~\ref{sec:intro} has 
its generating function 
$\displaystyle{g_{\psi}(z) = \frac{z^5+z^6}{1-z^4-z^5}}$ (see explanations below) and 
the convergence radius is $0.85667$. 
The mean length is $\displaystyle{N_z=\frac{5+6z-z^4-2z^5-z^6}{1+z-z^4-2z^5-z^6}}.$
So for a desired $N$ one has to find $z\in[0,0.85667)$ such that $5-N+(6-N)z+(N-1)z^4+(2N-2)z^5+(N-1)z^6=0$. 
For a mean length $N=15$, the polynomials is $-10-9z+14z^4+28z^5+14z^6$ and one has to instantiate $z$ with the root 
$0.78631$. One can also check that $z=0$ yields $N_z=5$: the corresponding Boltzmann sampler always outputs the smallest shape word which is $ABCFA$. 
%

To compute the generating function, it suffices to use the inductive 
definition:  
$g_a(z)=z$ for every $a\in\{A,\ldots,F\}$, $g_{DE\cup F}(z)=z^2+z$, and 
$g_{ABC(DE \cup F)}=z^4+z^5$. 
Then, $\displaystyle{g_{(ABC(DE \cup F))^+}(z)=g_{ABC(DE \cup F)}(z) \;  g_{(ABC(DE \cup F))^*}(z)= \frac{z^4+z^5}{1-z^4-z^5}}$. 
Finally we multiply by $z$ for the concatenation of $A$ to get the formula for $g_{\psi}$.
\end{example}




\subsection{Sampling Constrained Parameters in Shape Expressions}
\label{sec:params}
We explain the the 
{\em hit-and-run} algorithm that we use to sample 
parameter constraints in shape expressions.

Let $S \subseteq \R^{n}$ be a {\em bounded} {\em open} set. Let $\unitsphere=\{(x_1,\ldots,x_n)\mid \sum_{i=1}^n x_i^2=1\}$ be the set of n-dimensional unit vectors; 
every vector of 
$\unitsphere$  defines a direction. Algorithm~\ref{alg:hitrun} describes the hit-and-run method in its most abstract form and can be applied to 
non-linear constraints.
The procedure translates a point $X_k \in S$ into another point $X_{k+1} \in S$ by randomly choosing a direction $\Theta_{k}$ in 
 $\unitsphere$ and then randomly choosing the point $X_{k+1}$ uniformly distributed on the intersection $\Xi_{k}$ between the line crossing $X_{k}$ along the direction $\Theta_{k}$ and $S$. This procedure is repeated until some stopping criterion (based on, for instance, the sample number or computation budget) is met.
\begin{algorithm2e}
Initialize $X_{0} \in S$, set $k := 0$ and $\emph{stop} := 0$\;
\While{$\neg \emph{stop}$}{
Choose a random direction $\Theta_{k}$ in $\unitsphere$ with uniform distribution\;
Choose a random point $X_{k+1} \in \Xi_{k}$ with uniform distribution, where $\Xi_{k} = \{X_k + r\Theta_{k}~|~r \in \R\} \cap S$ \;
Set $k := k+1$ \;
$\emph{stop} := \emph{stopping\_criterion()}$\;
}
 \caption{Hit-and-run method $\textsf{hr}$.}
  \label{alg:hitrun}
\end{algorithm2e}
Algorithm~\ref{alg:hitrun} has two non-trivial steps: (1) the initial step of choosing $X_{0}$ in $S$, and (2) the line sampling step consisting of choosing a point in $\Xi_{k}$ with 
uniform distribution. 

\noindent {\bf Non-linear constraints.} The initialization step for arbitrary open bounded sets is done using 
optimization, such as {\em particle swarm optimization} (PSO)~\cite{488968}, an iterative
optimization method that tries in every step to improve a candidate solution with regard to a given measure of quality. 

\begin{figure}
\centering
\includegraphics[width=\textwidth]{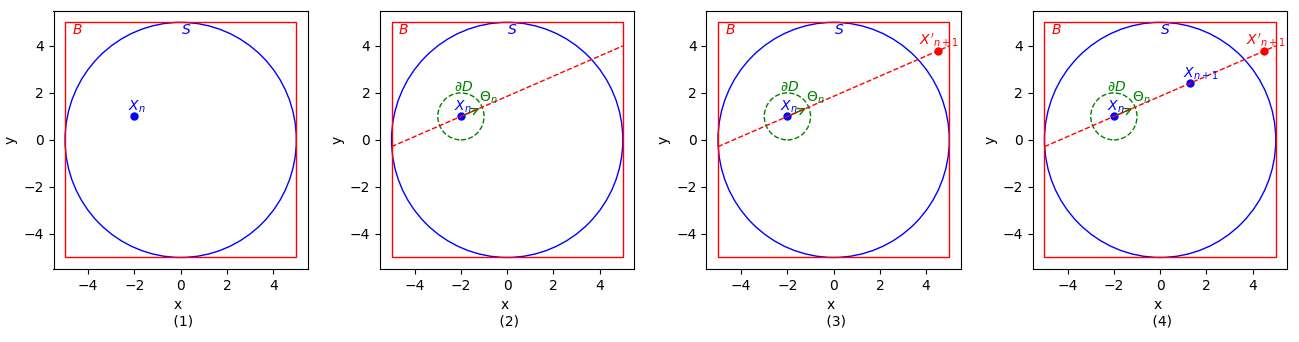}
\caption{An illustration of hit-and-run with the bounding hyperrectangle: (1) sample 
$X_{n}$ in $S$, (2) line segment $\tilde{\Xi_{n}}$ which is the intersection with $B$ of the line with random direction $\Theta_{n}$ passing through $X_{n}$, (3) rejected sample $X'_{n+1}$ on the line segment $\tilde{\Xi_{n}}$ 
(such that $X'_{n+1} \in B$ and $X'_{n+1} \not \in S$), and (4)
accepted sample $X_{n+1}$ on the line ($X_{n+1} \in B$, $X_{n+1} \in S$).}
\label{fig:hrexample}
\end{figure}

The line sampling step is done by first defining an $n$-dimensional hyperrectangle $B$ that encloses $S$. 
Next, we intersect the line 
crossing $X_k$ along the direction $\Theta_{k}$ with $B$. The resulting line segment, denoted by $\tilde{\Xi_{k}}$, contains points that 
can be inside or outside $S$. We then employ $1$-dimensional acceptance/rejection sampling on this line, until we get a point that is a member of $S$. Figure~\ref{fig:hrexample} depicts the main steps of this procedure. \\

\noindent {\bf Polynomial constraints.}
For the set $S$ defined by a set of polynomial constraints of degree $m$ in the form:
$\gamma = \bigwedge_{i=1}^{k} c^{i}_{0} + c^{i}_{1}x + c^{i}_{2}x^{2} + \ldots + c^{i}_{m}x^{m} > 0$, we can 
use a constraint solver~\cite{DBLP:journals/cca/JovanovicM12} to find an initial point that 
lies within $S$. 

\noindent {\bf Accelerated hit-and-run.} 
We can accelerate hit-and-run 
by using the so-called 
{\em shrinking} method~\cite{neal2003}, which 
works as follows: (1) let $r_{min} = min(\tilde{\Xi_{k}})$ and $r_{max} = max(\tilde{\Xi_{k}})$; (2) Choose a random point $r' \in (r_{min}, r_{max})$ with uniform distribution and let $X' = \Theta_{k}r' + X_{k}$; (3-a) if $X' \in S$, then $X_{n+1} = X'$; (3-b) if $X' \not \in S$, then: (3-b-I) if $r' < 0$, then $r_{min} = r'$; (3-b-II) if $r' \geq 0$, then $r_{max} = r'$; (3-b-III) goto step~(2).

Hit-and-run with shrinking has the same theoretical mixing time as without shrinking for convex sets, for non-convex sets it is not known. But empirically, the efficiency of hit-and-run with shrinking is superior~\cite{KiatsupaibulSZ11}. \\
~\\

\noindent {\bf Coordinate directions hit-and-run.}
The mixing time of hit-and-run can be further accelerated%
by using the coordinate directions hit-and-run (CDHR)~\cite{hitrun} variant. In CDHR, a different direction set $\Theta_{k}$ is used, namely the set of axis-aligned unit vectors. In essence, 
the new direction goes along one of the (randomly chosen) dimensions.
This introduces a trade-off between potentially more mixing steps but simpler computation of the intersection with the bounding box.
CDHR is also shown to converge to the uniform distribution asymptotically~\cite{hitrun}.
%
CDHR outperforms in practice the classical hit-and-run, 
especially in presence of parameters with different 
scales.

\noindent {\bf Asymptotically uniform distribution of noiseless signals for a fixed shape word.}
The statement (2) of Theorem~\ref{theorem:prop} about the asymptotically uniform distribution of sampled noiseless signals for a fixed shape word is guaranteed by the fact the hit-and-run algorithm defines a Markov chain on the set $S$ that has a uniform stationary distribution on $S$. This means that if we sample sufficiently long, the distribution of the sampled parameters converges to the uniform distribution on $S$. Concerning the mixing time of hit-and-run, if the set is convex, it is proven to be polynomial
~\cite{Lovasz1999}. If this set is non-convex, the mixing time is polynomial under some  condition of smoothness and curvature of the set~\cite{YadkoriBGM2017}. These results can be directly applied to our sampling procedure, depending on the constraints in the shape expressions of interest.   
\section{Case Studies and Experimental Results}
\label{sec:experimental}
We evaluate our sampling approach on several 
examples, including a visualization of an 
ECG specification and testing an aircraft elevator 
control system (AECS) in which we use SE to 
constraint inputs.
%
%
%
We implemented a
prototype sampler in Python 3.6.
We used the Pyswarm library for PSO\footnote{\url{https://pythonhosted.org/pyswarm/}}, and Z3\footnote{\url{https://github.com/Z3Prover/z3}}~\cite{DBLP:conf/tacas/MouraB08} for SMT solving.
All the experiments were carried out on a computer 
with the Intel Core I7-8650U processor with 4 cores, 
16 Gb RAM and Windows 10 Professional operating system.

\subsection{Visualisation of Shape Expressions}
\label{sec:vis}

In this section, we demonstrate the use 
of shape expression sampling to visualize and interpret specifications. 
We consider visualizations of two specifications: (1) the pulse specification with polynomial 
constraints and a complex regular 
expression structure, and (2) a heart 
beat specification inspired from ECG signals that has thin non-linear constraints. 

\begin{figure}
    \centering
    \includegraphics[width=\textwidth]{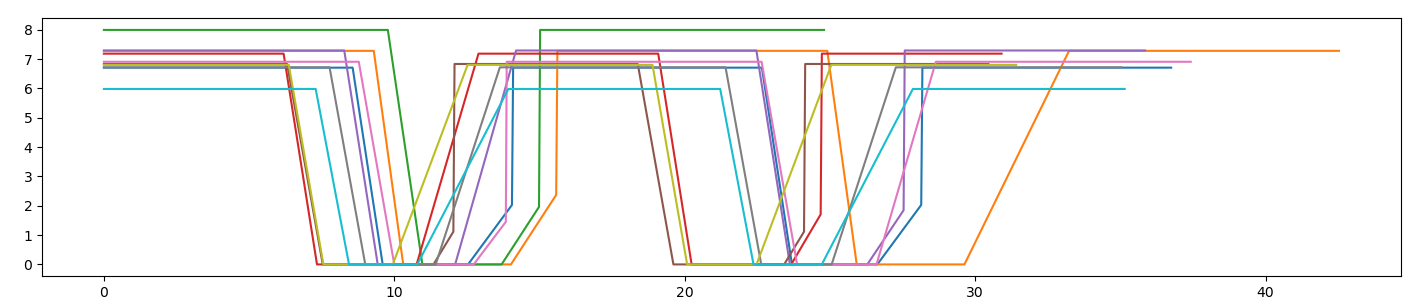}
    \caption{Visualizing the possible behaviors of the pulse shape expression with complex regular expressions structure.}
    \label{fig:bolz}
\end{figure}

Figure~\ref{fig:bolz} depicts $10$ samples of 
the $\specp$ behavior generated by our sampler. 
We observe both qualitative (number and 
shape of pulses) and quantitative (different parameters) variations of generated behaviors. We note that $\specp$ has 
equality constraints, which cannot be 
handled by hit-and-run. We relax 
equality constraints by introducing an 
$\epsilon$ approximation.

For the heart beat example, we are inspired by 
ECG signals from PhysioBank~\cite{physiobank}, a database
containing 549 medical records including ECG 
readings from 209 male and 81 female subjects, 
aged from 17 to 87. 
The database covers various diagnostic 
conditions related to cardio-vascular diseases.

In this experiment, we manually define an SE specification
for heart-beats inspired by the ECG from the database. This example illustrates visualisation 
of a specification that 
combines exponential and linear shapes. 
This non-linear aspect makes it 
interesting for exploring the visualization 
application. The heart-beat 
specification is formalized as follows: $\spech = A \cdot B \cdot C \cdot D \cdot E \cdot F \cdot G~:~\gamma$, where
\[
\small
\begin{array}{lclclcl}
A & = & \exp_x(a_{1},b_{1},c_{1},d_{1}) &\;\; & \gamma & = & a_1 = 0 \wedge b_{1} \in (0.008, 0.027) \wedge c_{1} \in (30, 32) \;\wedge \\
&&&&&& d_{1} \in (0.046, 0.047) \;\wedge \\ 
B & = & \exp_x(a_{2},b_{2},c_{2},d_{2}) &&&& a_2 \in (0.03, 0.1) \wedge b_{2} \in (0.08, 0.23) \wedge c_{2} \in (-35, -32) \;\wedge \\
&&&&&& d_{2} \in (0.101, 0.102) \wedge a_1 + b_1\cdot e^{c_1\cdot d_1} = a_2 + b_2 \;\wedge \\
C & = & \lin_x(a_{3},b_{3},d_{3}) &&&& a_3 \in (22, 30) \wedge b_{3} \in (-0.12, 0.01) \wedge d_{3} \in (0.03, 0.031) \;\wedge \\
&&&&&& a_2 + b_2\cdot e^{c_2\cdot d_2} = b_3 \;\wedge \\
D & = & \lin_x(a_{4},b_{4},d_{4}) &&&& a_4 \in (-50, -30) \wedge b_{4} \in (0.7,0.8) \wedge d_{4} \in (0.027, 0.028) \;\wedge \\
&&&&&& a_3\cdot d_3 + b_3 = b_4 \;\wedge \\
E & = & \lin_x(a_{5},b_{5},d_{5}) &&&& a_5 \in (20, 30) \wedge b_{5} \in (-0.4,-0.3) \wedge d_{5} \in (0.012, 0.013) \;\wedge \\
&&&&&& a_4\cdot d_4 + b_4 = b_5 \;\wedge \\
F & = & \exp_x(a_{6},b_{6},c_{6},d_{6}) &&&& a_6 \in (-0.05, 0.03) \wedge b_{6} \in (0.018, 0.043) \wedge c_{6} \in (8, 9) \;\wedge \\
&&&&&& d_{6} \in (0.15, 0.1525) \wedge a_5\cdot d_5 + b_5 = a_6 + b_6 \;\wedge \\
G & = & \exp_x(a_{7},b_{7},c_{7},d_{7}) &&&& a_7 \in (-0.02, 0.123) \wedge b_{7} \in (0.0395, 0.0415) \;\wedge \\
&&&&&& c_{7} \in (-35, -34) \wedge d_{7} \in (0.046, 0.047) \;\wedge \\
&&&&&& a_6 + b_6\cdot e^{c_6\cdot d_6} = a_7 + b_7.
\end{array}
\]

\begin{figure}
\centering
\begin{subfigure}{.45\linewidth}
  \centering
\includegraphics[width=\linewidth]{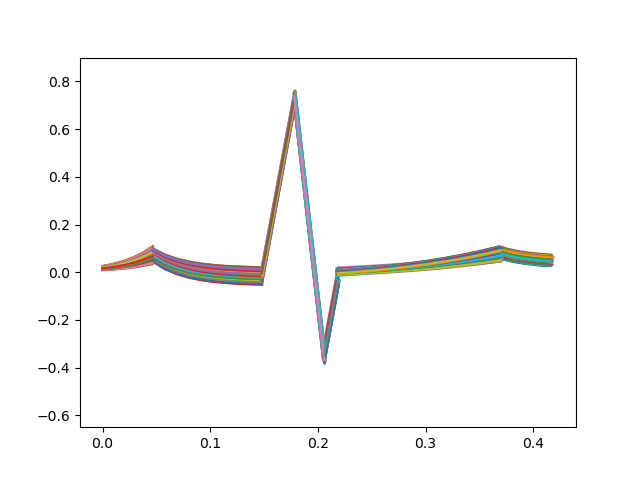}
  \caption{classical hit-and-run}
  \label{fig:sub1}
\end{subfigure}%
\begin{subfigure}{.45\linewidth}
  \centering
\includegraphics[width=\linewidth]{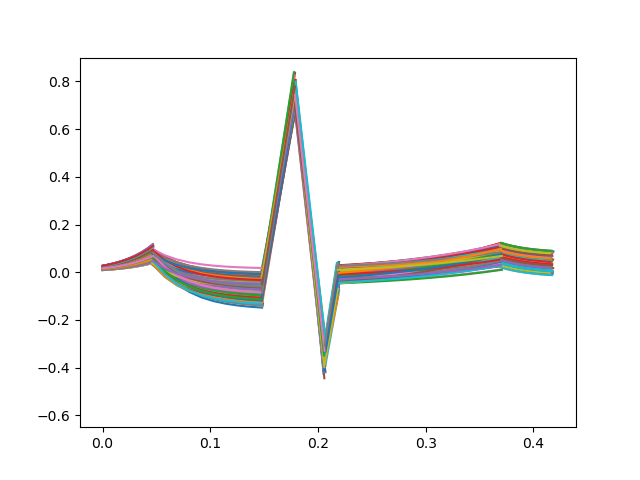}
  \caption{coordinate direction hit-and-run}
  \label{fig:sub2}
\end{subfigure}
\caption{Behaviors generated from $\spech$.}
\label{fig:ekg}
\end{figure}

Due to the non-linearity of the constraints, 
we can only employ the hit-and-run variant 
that uses optimization to find the initial point satisfying 
the constraints. We also remark that the 
constraints of the specification are thin. We compare the performance between the classical and the coordinate direction hit-and-run to generate samples and show the 
results in Figure~\ref{fig:ekg}. 
In each case we generated $100$ behaviors using 
$1000$ mixing steps. The classical hit-and-run 
algorithm required $193s$ to generate these behaviors 
with $0.36\%$ acceptance rate, while the coordinate 
direction version of the procedure required 
$196s$ with $0.14\%$ acceptance rate.
We can observe 
that the coordinate direction variant of the 
algorithm achieves much better mixing with the 
same number of sampled behaviors.

\subsection{Testing an Aircraft Elevator Control System with Shape Expressions and Signal Temporal Logic}
\label{sec:falsification}

In this section, we demonstrate our approach of sampling shape expressions for systematic testing of 
the Aircraft Elevator Control System (AECS)~\cite{mosterman-fdir}, a CPS Simulink 
model that illustrates model-based design of 
Fault Detection, Isolation and Recovery (FDIR) 
in control systems by using actuation redundancy.


AECS has two elevators, 
on the left and one on the right side. Both elevators 
are equipped with two hydraulic actuators. 
Each actuator can position the elevator. 
%
Formal verification of CPS models such as AECS comes 
at a prohibitive cost. Even the medium-size AECS model 
represents a real challenge 
for exhaustive verification, with $426$ real-valued, 
Boolean and enumerated (state) variables. This 
situation favors  more pragmatic {\em simulation-based testing}. This a priori incomplete verification 
activity can be made more 
principled by integrating formal specifications to 
express desired properties.

{\em Signal Temporal Logic} (STL)~\cite{DBLP:conf/formats/MalerN04} 
is a popular declarative language for specifying CPS requirements. 
STL admits {\em quantitative} semantics~\cite{fainekos-robust,DBLP:conf/formats/DonzeM10} 
that allows defining the {\em robustness degree} $\rho(w, \psi)$ of a STL specification $\psi$ to a (simulation) trace $w$, that is it measures how far $w$ is from satisfying or violating 
$\psi$. {\em Falsification-based testing}~\cite{DBLP:conf/hybrid/NghiemSFIGP10} 
is a popular approach that uses the robustness degree 
to steer the CPS model to the specification violation.

A typical system-level property of AECS requires that the intended position of the aircraft given by the pilot 
shall be achieved within a predefined time window 
and with a predetermined accuracy. 
This can be captured with several requirements. One 
requirement states that when the Pilot Command $\emph{pc}$ has a derivative higher than some 
threshold $m$, the left actuator position $\emph{lep}$ measured by the sensor must 
stabilize (become at most $n$ units away 
from the pilot command input) within 
$T+t$ time units (for some $t$ and $T$). 
We formalize this requirement as the following 
STL specification:

\begin{equation}\label{stlFormula}
\stlspec \equiv \Box ((\emph{pc}' \geq m) \rightarrow \Diamond_{[0,T]} \Box_{[0,t]} (|\emph{pc} - \emph{lep}| \leq n)).
\end{equation}
\noindent with $m=10$, $T=300$, $t=50$ and $n=0.1$.

We evaluate the following scenario - 
principled testing of a system-under-test (AECS model) 
guided by its expected properties formalized in 
the STL specification language (formula $\stlspec$) 
under the constrained input space given by a 
shape expression (specification $\specp$). In 
particular, we show how to use our shape expression 
sampling approach to improve: (1) simulation-based 
{\em sensitivity analysis} of the model, and (2) 
falsification-testing of the model.

\begin{figure}
        \centering
    \includegraphics[width=\textwidth]{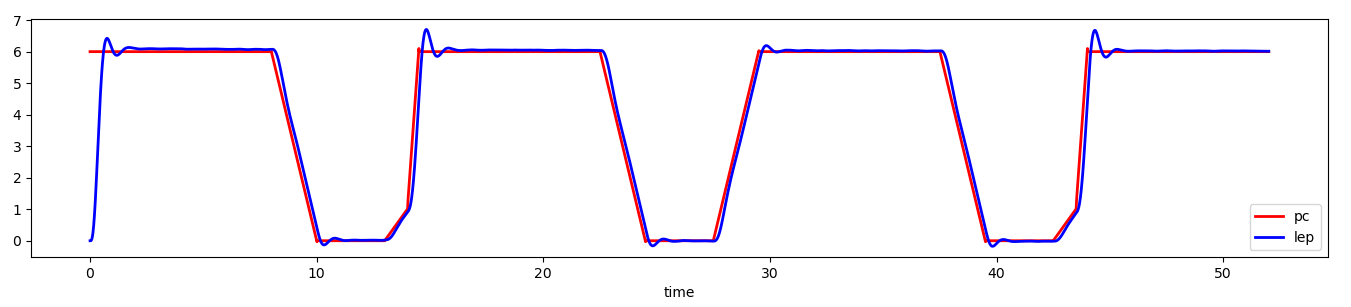}
\caption{The pulse train as pilot command input to AECS (red) and the resulting left elevator position (blue).}
\label{fig:pulse-aecs}
\end{figure}

We assume that the input 
(pilot command) signal is constrained with 
the shape expression $\specp$ from the 
introduction. Figure~\ref{fig:pulse-aecs} 
depicts the pilot command input (in red) from 
Figure~\ref{fig:pulse}, and the resulting 
left elevator position output (in blue) obtained 
by executing the model in Simulink.

\begin{figure}
        \centering
    \includegraphics[width=\textwidth]{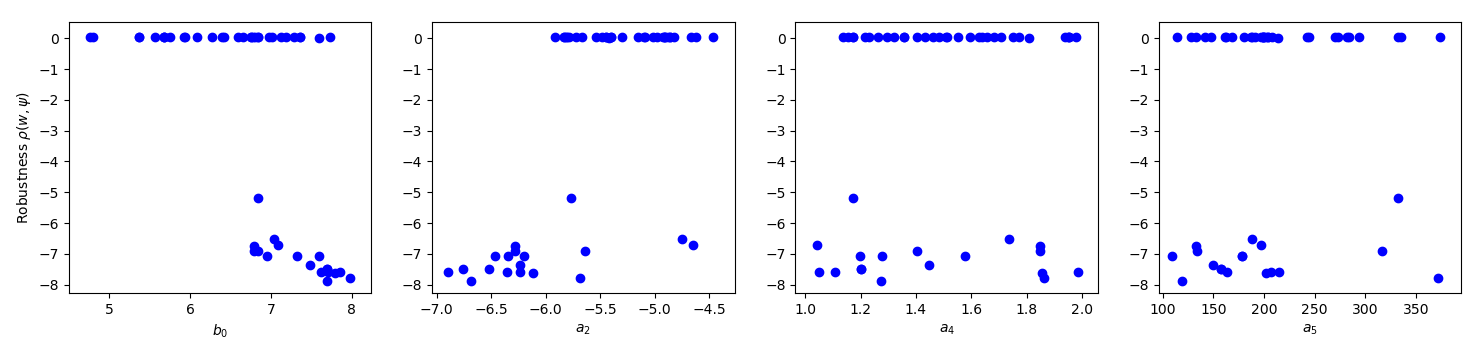}
\caption{Sensitivity with respect to robustness for parameters $b_{1}$, $a_{2}$, $a_4$ and $a_5$.}
\label{fig:sens}
\end{figure}

In the first experiment, we generate $50$ 
sampled behaviors from the shape expression 
$\specp$. To improve the uniformity of the 
samples' distribution, we adopt two 
orthogonal approaches: (1) hit-and-run 
with coordinate direction, and (2) 
accepting on average only every 
$100^{\emph{th}}$ sample. We simulate 
the AECS model with each of the generated 
input behaviors and compute robustness 
with respect to the STL requirement $\stlspec$. 
We then study how individual parameters 
from $\specp$ affect the robustness of $\stlspec$.
Figure~\ref{fig:sens} shows the results 
for parameters $b_{0}$, $a_{2}$, $a_{4}$ 
and $a_{5}$. 
We can see that the 
offset $b_{2}$ at which the pulse starts  
and the steepness of the slope $a_{2}$ of 
the second segment in the pulse significantly 
affect the resulting robustness. 
The individual role of parameters $a_{4}$ and 
$a_{5}$ is much less clear.

In the next experiment, we consider the falsification problem for the formula 
$\stlspec$ with constrained inputs: 
$$
\min_{w} \rho(w~||~F(w), \stlspec) \text{ subject to } w \in S,
$$
\noindent where $S$ is the set of constrained inputs, 
$w~||~F(w)$ is the combined input 
($w$) and output ($F(w)$) behavior of the AECS 
model, $\stlspec$ is an STL formula and $\rho$ is the STL
robustness function.

\begin{figure}
    \centering
    \includegraphics[width=\textwidth]{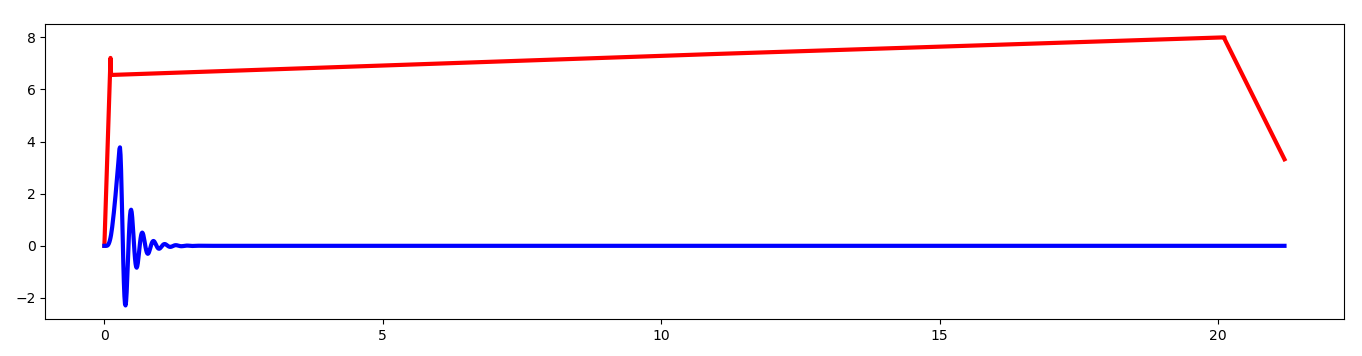}
    \caption{Input behavior and its associated output witnessing the violation of $\stlspec$.}
    \label{fig:witpso}
\end{figure}

We first study the problem where 
$S$ is a simple hyper-rectangle, i.e. every 
input is bounded by an interval. We define 
the input behavior with $5$ control points, 
each having a timestamp and a value. 
The value of 
each control point is in $[0,8]$ and the 
distance between consecutive control points is 
in $[0.1, 10]$. This scenario reflects 
the common use case  
of falsification testing, where inputs are 
represented as sequences of control 
points. We solve the above optimization 
problem with Pyswarm PSO library 
with swarm size $10$, the maximum number of 
iterations for the swarm to search $10$, 
particle velocity scaling factor $0.5$, 
and the scaling factor to search away from the 
particle's best position $0.5$. After $20$ 
simulations and $94s$, the 
procedure finds an input sequence $u$ that 
steers the model to the violation of $\stlspec$ with negative robustness 
of $-6.57$. The input behavior, and its 
associated output are shown in Figure~\ref{fig:witpso}. 
We can appreciate that the input behavior does not 
correspond to a realistic pilot command.

We next consider the scenario where 
$S$ is defined by the pulse-train 
shape expression $\specp$. 
We can solve in theory this optimization problem 
in a straightforward manner that does not 
require shape expression sampling. 
Pyswarm PSO library allows restricting 
inputs with arbitrary 
constraints. In this 
case, we encode the input as a set of parameters 
that uniquely represent the given shape expression. 
We then enrich the PSO procedure with a function 
which evaluates the generated input against 
the constraints from $\specp$. PSO 
uses a rejection procedure that only applies 
inputs satisfying the constraints from $\specp$.
We ran the PSO procedure with the same parameters 
as above. The algorithm could not find an 
input that leads to the property violation 
after $353s$ and $102$ model simulations. 
We suspect that this straightforward procedure 
fails because it uses a trivial rejection sampling 
strategy. Given that the input space is both 
many-dimensional and ``thin'', the procedure 
fails to find a point within the 
constraints.

\begin{figure}
    \centering
    \includegraphics[width=\textwidth]{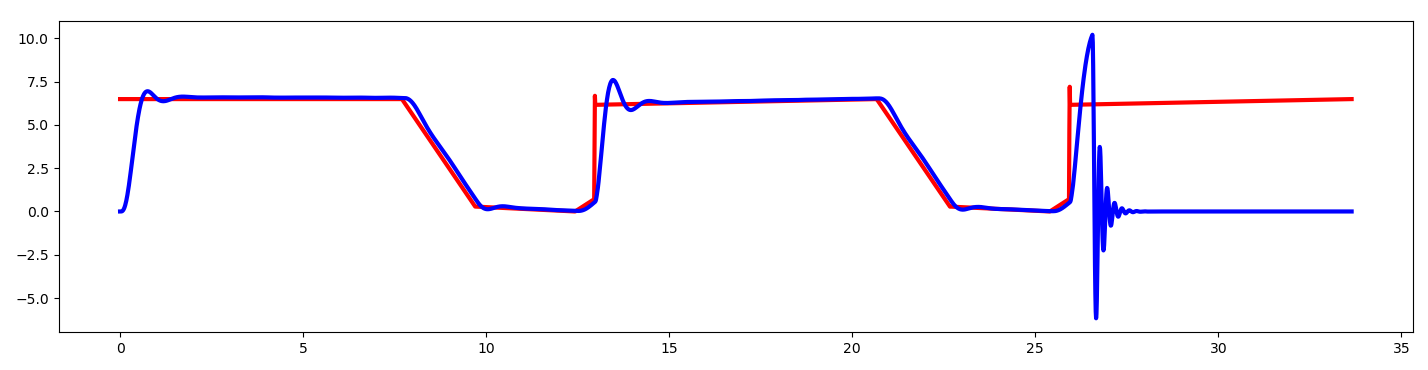}
    \caption{Pulse train that steers AECS to specification violation.}
    \label{fig:witness}
\end{figure}

Finally, we implement a procedure 
for falsification-testing with inputs constrained 
by a shape expression that incorporates 
the hit-and-run method. In every iteration, 
it uses hit-and-run to generate 
a new candidate input that 
is guaranteed to satisfy the shape expression. 
This input behavior is used to simulate the 
model, and the outcome is checked for robustness. 
The procedure updates the current best candidate 
only if it improves the robustness. In this 
experiment, we were able to find the input behavior that 
both satisfies the shape expression $\specp$ and 
steers the model to violate $\stlspec$ in $10$ iterations
after $32s$. Figure~\ref{fig:witness} shows the input and
its associated output behavior. The reader can 
appreciate the quality of the input with respect 
to the unconstrained input from Figure~\ref{fig:witpso}.

\section{Related Work}
\label{sec:related}

The hit-and-run sampling algorithm was first introduced in~\cite{hitrun}, where the author showed that the algorithm 
generates samples of a bounded set that are asymptotically uniform. The shrinking method that enables better 
convergence in the rejection sampling of a line was introduced in~\cite{neal2003} and applied to hit-and-run 
in~\cite{KiatsupaibulSZ11}. The major steps of these methods, formulated in these papers in an abstract manner, are efficiently concretized and adapted in our work for sampling points in (possibly non-convex) semi-algebraic sets, which together with the Boltzmann sampling allows systematically generating signals from shape expressions with asymptotically uniform distribution. 

Hit-and-run was previously exploited by the formal methods community for falsification of hybrid systems~\cite{DBLP:conf/hybrid/NghiemSFIGP10}, the focus of which is on using quantitative semantics of temporal specifications to guide the system-under-test towards the violation of the property, where hit-and-run method is proposed when exploring convex input domains. Note that the input domains in this work are subsets of $\R^n$. More recent work on falsification of hybrid and cyber-physical systems extends inputs to sets of signals described by piecewise functions encoded using a finite number of bounded parameters. The literature on CPS falsification and testing has become vast, we include here only some most recent papers~\cite{DreossiDS19,ErnstSZH19,ZhangAH20,BassetBD2019,BarbotBDDKY20} and more references can be found therein as well as in the papers and websites related to the major state-of-the-art tools S-TaLiRo~\cite{STaliro}
and Breach~\cite{Donze20}. Compared to this approach, our work can deal with more general signals described by complex constraints involving both time and value domains. Along this line, the closest work to ours is~\cite{BassetBD2019} where the temporal patterns of input signal spaces are constrained by timed automata which entails particular polytopic constraints that are nevertheless simpler than the constraints yielded by shape expressions. 
Uniform sampling of traces satisfying given constraints has been considered in {\em constrained random verification}, for different specification and modeling formalisms, such as uniform and nearly-uniform sampling satisfying assignments of Boolean formulas~\cite{DBLP:conf/cav/ChakrabortyMV13,ChakrabortyFSV2015,GuptaSRM2019}, uniformly sampling traces of networks of automata~\cite{DBLP:conf/concur/BassetMS17}, timed automata~\cite{DBLP:conf/qest/BarbotBBK16}, transition systems~\cite{ChakrabortySV2020}. Monte-Carlo LTL model-checking introduced in~\cite{GrosuSmolka2005} was enhanced with a uniform version~\cite{OudinetDGLP2011}. Uniform sampling can be used as a means of providing theoretical guarantees for random testing of distributed programs~\cite{OzkanMNBW2018}. Note that our sampling method uses Boltzmann sampling, as in~\cite{DBLP:conf/concur/BassetMS17}, for the discrete part (namely shape words) of shape expressions. However, unlike all the above-mentioned work, our work can deal with both temporal and spatial constraints of cyber-physical signals. As mentioned in the introduction, generating cyber-physical 
behaviors from STL was 
proposed in~\cite{PrabhakarLK18},
where STL specifications are encoded as linear 
arithmetic real formulas and an SMT solver is used 
to find satisfying assignments in the form of 
signal behaviors. In contrast to our approach, 
this work does not aim at sampling specifications 
according to any specific distribution.

In the spirit of using randomness for exploration, statistical model-checking SMC, which is based on sampling behaviors of a system model and using statistical inference to estimate the probability that the system satisfies a property, has been applied for CPS (see some recent surveys~\cite{LarsenLegay2014,smc-overview,AghaP18,Legay2019,Pappagallo2020} which include descriptions of the existing SMC tools). Unlike the concern of this approach involving rare events which is addressed by techniques such as importance sampling, cross-entropy methods, trace uniformity requirements have not been explored. They can be additionally considered to provide coverage guarantees especially useful when there is no information about falsifying behaviors as well as when no falsifying behaviors are revealed by the validation process.





\section{Conclusions and Future Work}
\label{sec:conclusion}

We presented an approach for generating random 
behaviors from shape expressions that are asymptotically 
uniform. It combines algorithms that 
were originally studied in the combinatoric community
(Boltzmann sampling) and in operation research (hit-and-run 
sampling), exploring different variants of these 
procedures. We thus provided an effective approach for 
sampling sophisticated temporal specifications and demonstrated 
its usefulness for the visualisation and interpretation of 
formal specifications, as well as for systematic 
CPS testing.

We plan to study sampling of 
shape expressions in the presence of equality constraints. 
In this work, equality constraints were replaced by ``thin''
open constraints which significantly 
increase the mixing times needed to achieve almost-uniformity 
of samples. While we tackled this problem with 
coordinate direction variant of hit-and-run, 
there is room for improving the methods that
would allow faster convergence. In this paper, 
we used shape expressions together with the hit-and-run 
algorithm to narrow the input exploration space. We plan to improve the interaction between 
optimization engines and the shape expressions sampler to 
improve falsification testing.


\bibliographystyle{plain}
\bibliography{main}

\appendix
\appendix

\section{Evaluation of Hit-and-Run}
\label{sec:eval}

We evaluate different 
hit-and-run variants presented in Section~\ref{sec:sampling}.
In our experiments, classical rejection sampling, denoted by 
$\rs$, is used as our baseline. 

Our specification is defined by the $n$-dimensional hyper-ring:
$\chi = x_{1}^{2} + \ldots + x_{n}^{2} < c_1^{2} \;\; \wedge \;\;
 x_{1}^{2} + \ldots + x_{n}^{2} > c_2^{2}$ where $c_1 > c_{2}$, and the hyper-box constraint $\psi$ embodying the hyper-ring: $\psi = \bigwedge_{i=1}^{n} x_{i} > - c \;\; \wedge \;\; x_{i} < c$ where $c \geq c_1$. This hyper-ring specification has several 
interesting characteristics: (1) it is polynomial, (2) it is non-convex, 
(3) it can be instantiated to an arbitrary number of dimensions, and 
(4) it can be made arbitrarily ``thin''.

\begin{figure}
    \centering
    \includegraphics[width=\linewidth]{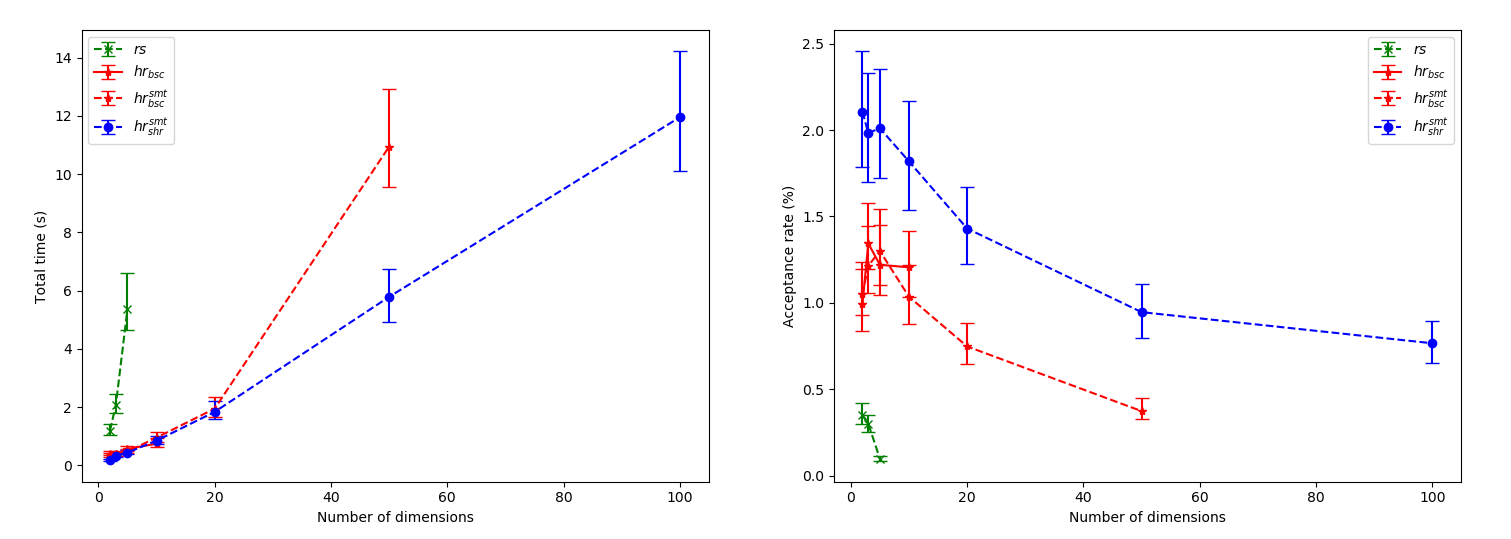}
    \caption{Performance of hit-and-run in number of dimensions: (1) computation time (left) and 
    (2) acceptance rate (right).}
    \label{fig:dim}
\end{figure}

We first study hit-and-run variants 
with respect to the number of dimensions. 
For this experiment, we generate $100$ samples from 
the set defined by $\chi \wedge \psi$, where $c = c_{1} = 1$ and $c_{2} = 0$.
We vary the number of dimensions in the specification and 
repeat each experiment $5$ times. Figure~\ref{fig:dim} summarizes 
the results. 

We first remark that classical rejection sampling does not scale beyond 
$3$ dimensions. Also, the hit-and-run variants that 
use search fail to find an initial sample in more than $10$ 
dimensions. The use of the SMT solver helps to efficiently find 
an initial sample and enables scaling to $50$ dimensions. 
We finally observe that the shrinking mechanism allows halving 
the computation time (and sampling $100$ points in a $100$-dimensional ball in $12s$) 
and considerably improving the acceptance rates, clearly outperforming 
the ``vanilla'' version of hit-and-run.

\begin{figure}
    \centering
    \includegraphics[width=\textwidth]{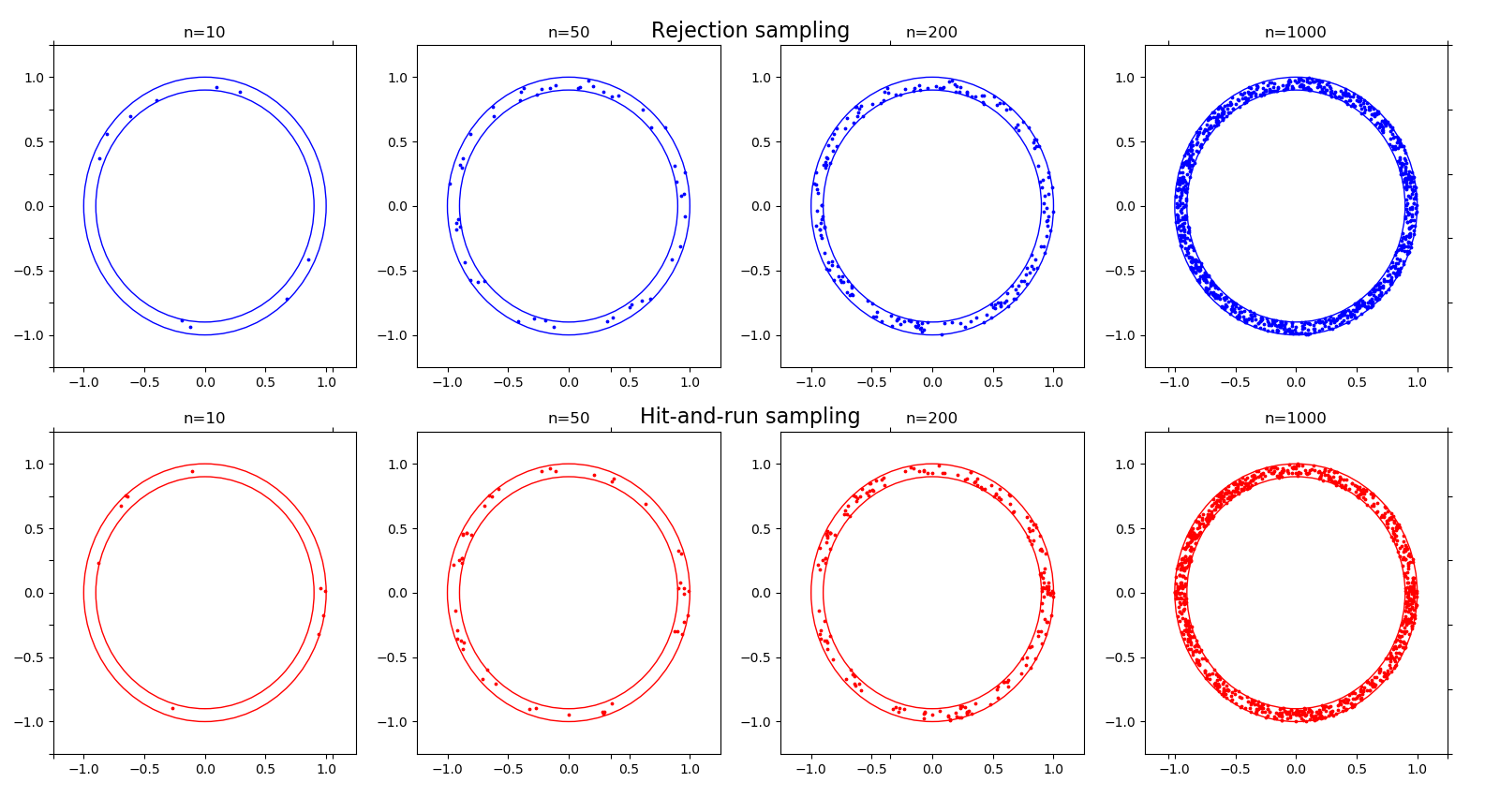}
    \caption{Evolution of sample distribution generated by hit-and-run compared 
    to that by rejection sampling.}
    \label{fig:dist}
\end{figure}

In the next experiment, 
we study the distribution of samples that it generates for 
hyper-rings of various thickness. 
In Figure~\ref{fig:dist}, we visually compare the evolution of 
the sample distributions generated from a constraint defining a 
$2$-dimensional ring with $c_1 =1$ and $c_2=0.9$ 
by rejection sampling and by hit-and-run, where we take 
snapshots after generating $10$, $50$, $200$ and $1000$ samples. 

We conduct two more experiments that evaluate the acceptance rate 
of hit-and-run with respect to: (1) 
the thickness of the hyper-ring, and (2) the size of the 
bounding hyper-box. In these experiments, we consider a $3$-dimensional 
hyper-ring, in which we sample $1000$ times. We repeat each experiment 
$5$ times.

\begin{figure}
    \centering
    \includegraphics[width=\textwidth]{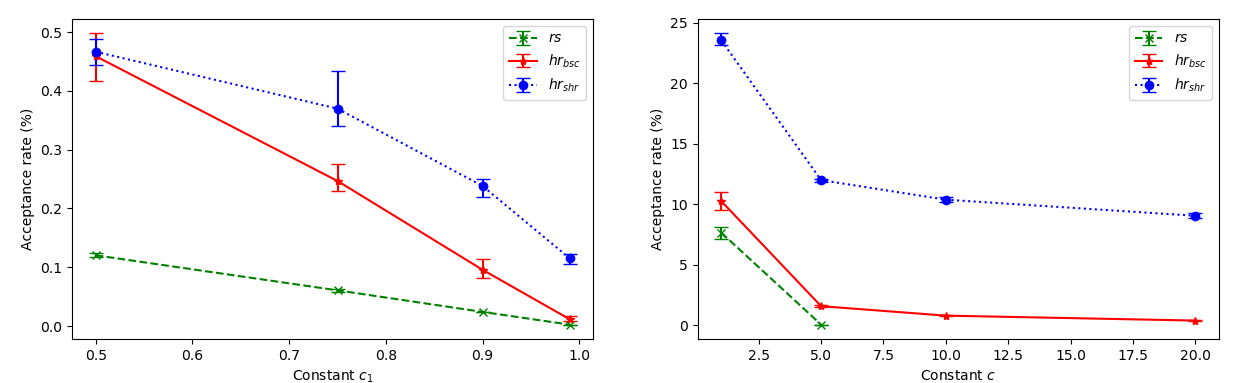}
    \caption{Acceptance rate with respect to the thickness of the 
    ring (left) and to the size of the bounding box (right).}
    \label{fig:rbs}
\end{figure}

We vary the hyper-ring thickness 
by fixing the constants $c$ and $c_{1}$ to $1$, and by varying the value of  
 $c_2$ among $0.5$, $0.75$, $0.9$ and $0.99$. 
Figure~\ref{fig:rbs} (left) summarizes the results. We can first 
observe that the thinner the hyper-ring is, the closer the 
acceptance rate of the basic hit-and-run algorithm comes to 
the standard rejection sampling. On the other hand, the shrinking 
mechanism allows to 
retain reasonable acceptance rates even for quite thin constraints 
(more than $10\%$ acceptance rate when $c_{1} = 1$ and $c_2 = 0.99$). 

We finally fix the constants 
$c_{1}$ to $1$ and $c_{2}$ to $0.9$ and vary the size of hyperrectangles 
(by changing the value of $c$) among $1$, $5$, $10$ and $20$. The results of this 
experiment are summarized in Figure~\ref{fig:rbs} (right). We first 
observe that acceptance rate of the rejection sampling procedure when
$c=5$ is only around $0.06\%$. We can also observe a similar trend 
of the acceptance rate dropping for the two hit-and-run procedures 
when the bounding box becomes large. We notice that the hit-and-run 
method with shrinking does perform better by an order of magnitude 
and maintains an acceptance rate above $10\%$ when $c=20$.

\end{document}